\documentclass[twocolumn,prd,a4paper,showpacs,amssymb,amsmath,nofootinbib,preprintnumbers]{revtex4}

\usepackage{graphicx}
\usepackage{dcolumn}% Align table columns on decimal point
\usepackage{bm}% bold math

\begin{document}

\def \d {{\rm d}}
\def \e {{\epsilon}}

\newcommand{\be}{\begin{equation}}
\newcommand{\ee}{\end{equation}}

\newcommand{\beqn}{\begin{eqnarray}}
\newcommand{\eeqn}{\end{eqnarray}}
\newcommand{\AdS}{anti--de~Sitter }
\newcommand{\AAdS}{\mbox{(anti--)}de~Sitter }
\newcommand{\AAN}{\mbox{(anti--)}Nariai }
\newcommand{\AS}{Aichelburg-Sexl }
\newcommand{\pa}{\partial}
\newcommand{\pp}{{\it pp\,}-}
\newcommand{\ba}{\begin{array}}
\newcommand{\ea}{\end{array}}

\title{Ultrarelativistic boost of the black ring}

\author{Marcello Ortaggio}
 \email{marcello.ortaggio@comune.re.it}
 \altaffiliation[also affiliated to ]{INFN (Rome)}
\author{Pavel Krtou\v{s}}
 \email{Pavel.Krtous@mff.cuni.cz} 
\author{Ji\v{r}\'{\i} Podolsk\'y}
 \email{Jiri.Podolsky@mff.cuni.cz}
 \affiliation{Institute of Theoretical Physics, Faculty of Mathematics and Physics, Charles University in Prague, \\
  V Hole\v{s}ovi\v{c}k\'{a}ch 2, 180 00 Prague 8, Czech Republic}

%\preprint{\tt arXiv:gr-qc/0503026}

\date{\today}

\begin{abstract}

We investigate the ultrarelativistic boost of the five-dimensional Emparan-Reall non-rotating black ring. Following the classical method of Aichelburg and Sexl, we determine the gravitational field generated by a black ring moving ``with the speed of light'' in an arbitrary direction. In particular, we study in detail two different boosts along axes orthogonal and parallel to the plane of the ring circle, respectively. In both cases, after the limit one obtains a five-dimensional impulsive \pp wave propagating in Minkowski spacetime. The curvature singularity of the original static spacetime becomes a singular source  within the wave front, in the shape of a ring or a rod according to the direction of the boost. In the case of an orthogonal boost, the wave front contains also a remnant of the original disk-shaped membrane as a component of the Ricci tensor (which is everywhere else vanishing). We also analyze the asymptotic properties of the boosted black ring at large spatial distances from the singularity, and its behaviour near the sources. In the limit when the singularity shrinks to a point, one recovers the well known five-dimensional analogue of the \AS ``monopole'' solution. 

\end{abstract}


\maketitle

\section{Introduction}

\label{sec_introduction}

The study of black holes in higher dimensions has been for a long time motivated by unified theories, in particular string theory \cite{MyePer86}. In the past few years, extra-dimension models of TeV gravity have raised further interest in view of possible black hole production at colliders \cite{BanFis99,EmpHorMye00prl,GidTho02,DimLan01}.
According to \cite{DratHo85npb,tHooft87}, in semi-classical investigations of such high energy phenomena one can represent the incoming states with black hole metrics boosted ``to the speed of light''. In the case of the four-dimensional Schwarzschild black hole, the corresponding ultrarelativistic gravitational field is described by the \AS impulsive \pp wave \cite{AicSex71}. In the spirit of \cite{BanFis99,EmpHorMye00prl,GidTho02,DimLan01}, however, one clearly needs to consider higher dimensional settings. Indeed, the boosting technique of \cite{AicSex71} has been already applied to static (charged) black holes in higher dimensions \cite{LouSan90} (and straightforwardly extended to the $D\ge 4$ Schwarzschild black hole in an external magnetic field \cite{Ortaggio05}). Recent analyses of black hole production in high energy collisions \cite{EarGid02,KohVen02,YosNam02} thus employed the \AS solution (or other impulsive waves) in $D\ge 4$ spacetime dimensions (see \cite{GidRyc04} for a subtler discussion). The very recent work \cite{Yoshino05} studied the more elaborate ultrarelativistic limit of the Myers-Perry solution \cite{MyePer86} (a generalization of the rotating Kerr metric to arbitrary dimensions). 

In fact, one of the most remarkable feature of General Relativity in $D>4$ is the non-uniqueness of the Myers-Perry spherical black holes. In five-dimensional vacuum gravity, there exist also asymptotically flat rotating black rings with an event horizon of topology $S^1\times S^2$ \cite{EmpRea02prl}. It is our purpose to investigate the gravitational field generated by such rings when they move at the speed of light, in the sense of the \AS limit. In the present paper we will be focusing on the special subcase of zero angular momentum, i.e. on the static black rings found in \cite{EmpRea02prd}. We shall consider spinning rings in a separate subsequent work \cite{OrtKrtPod05_2}. 

The structure of the paper is as follows. In Sec.~\ref{sec_ring} we briefly describe the static black ring of \cite{EmpRea02prd}, which we intend to Lorentz-boost subsequently. In Sec.~\ref{sec_splitting} we split the corresponding line element into flat space plus a term that becomes ``small'' at asymptotic infinity. We also introduce (asymptotically) cartesian coordinates useful for performing the ultrarelativistic boost. Secs.~\ref{sec_orthogonal}--\ref{sec_general} contain our main results. In Secs.~\ref{sec_orthogonal} and \ref{sec_parallel} we explicitly calculate the metric of the black ring boosted along a direction orthogonal and parallel to the plane of the ring circle, respectively. This leads to two different impulsive \pp waves that naturally ``recall'' the original curvature and conical singularities of the static ring (in a sense to be made clear later). We analyze several specific properties of such solutions, in particular the Ricci and Weyl tensors, and asymptotic expansions far from and close to the singularities, and near geometrically privileged axes and planes. In Sec.~\ref{sec_general} we briefly discuss a boost along an arbitrary direction. We again obtain an impulsive \pp wave, whose singular source is described by an ellipse. Our final remarks are presented in Sec.~\ref{sec_conclusions}. Appendix~\ref{app_integrals} summarizes the definitions and properties of the complete elliptic integrals employed in Secs.~\ref{sec_orthogonal} and~\ref{sec_parallel}, whereas Appendix~\ref{app_weyl} provides the explicit tetrad components of the Weyl tensor in the case of the orthogonal boost. 

\section{The static black ring}

\label{sec_ring}

In this section we briefly summarize the basic properties of the static black ring, referring to \cite{EmpRea02prd,Emparan04} for further details. In the coordinates of \cite{Emparan04},\footnote{More precisely, one has to multiply $F(\zeta)$ by $(1-\lambda)$, and to divide $\psi$ and $\phi$ by $\sqrt{1-\lambda}$ to obtain the corresponding quantities of \cite{Emparan04}. The original notation of \cite{EmpRea02prd} is recovered with the transformations $y=(y'-\lambda)/(1-\lambda y')$, $x=(x'-\lambda)/(1-\lambda x')$, $\psi=\psi'/\sqrt{1+\lambda}$, $\phi=\phi'/\sqrt{1+\lambda}$, $L^2=(1-\lambda^2)/A^2$.} the line element reads
\beqn
 \hspace{-1cm} \d s^2 & = & -\frac{F(y)}{F(x)}\d t^2+\frac{L^2}{(x-y)^2}F(x)\bigg[(y^2-1)\d\psi^2 \nonumber \label{ring} \\
  & & \hspace{-.2cm} {}+\frac{1}{F(y)}\frac{\d y^2}{y^2-1}+\frac{1}{F(x)}\frac{\d 
      x^2}{1-x^2}+(1-x^2)\d\phi^2\bigg] ,
\eeqn
where
\be
 F(\zeta)=\frac{1+\lambda\zeta}{1-\lambda} , \qquad 0\le\lambda<1 .
\ee
The parameter $\lambda$ is dimensionless, and for $\lambda=0$ [i.e., $F=1$] the spacetime (\ref{ring}) is flat. The constant $L>0$ represents a length related to the radius of the ``central circle'' of the ring. For a physical interpretation of the spacetime (\ref{ring}), we take
\be
 y\in(-\infty,-1] , \qquad x\in[-1,+1] ,
\ee
and $\psi$ and $\phi$ as periodic angular coordinates (see below). Now, $y$ is an ``area coordinate'' that, loosely speaking, parametrizes ``distances'' from the ring circle. Surfaces of constant $y$ have topology $S^1\times S^2$, and area which is monotonically growing with $y$. The coordinate $\psi$ runs along the $S^1$ factor, whereas $(x,\phi)$ parametrize $S^2$ (see \cite{EmpRea02prd,Emparan04} for illustrative pictures). At $y\to-\infty$ the spacetime has a curvature singularity, $y=-1/\lambda$ is a horizon of topology $S^1\times S^2$, and spatial infinity corresponds to $x,y\to -1$. To avoid conical singularities at the axes $x=-1$ and $y=-1$, the angular coordinates must have the standard periodicity 
\be
 \Delta\phi=2\pi=\Delta\psi .
 \label{range}
\ee
With this choice, however, there is a conical singularity at $x=+1$. This describes a disk-shaped membrane (with an excess angle) inside the ring which prevents the ring from collapsing under its self-gravity.\footnote{Alternatively, one can require regularity at $x=+1$ and place the conical singularity at $x=-1$, i.e., outside the ring \cite{EmpRea02prd}; we will not consider this case because the singularity would extend to infinity. There is no way to achieve regularity at both $x=-1$ and $x=+1$, unless the ring rotates \cite{EmpRea02prl}.} 
Nevertheless, the spacetime (\ref{ring}) is asymptotically flat \cite{EmpRea02prd}, and the black ring has mass
\be
 M=\frac{3\pi L^2}{4}\frac{\lambda}{1-\lambda} .
 \label{mass}
\ee

Except on the disk membrane at $x=+1$, the metric (\ref{ring}) is a vacuum solution. It clearly admits three commuting orthogonal Killing vector fields $\pa_t, \pa_\psi, \pa_\phi$ and, in fact, it belongs to the generalized Weyl class of \cite{EmpRea02prd}. Interestingly, it has been proven recently \cite{PraPra05} that vacuum black rings (with or without rotation) differ from the five-dimensional Myers-Perry black holes not only in the horizon topology, but also
in the algebraic type of the Weyl tensor: black holes are of type D, whereas black rings are of the more general type I$_i$ (type II on the horizon), according to the higher dimensional classification of  \cite{Coleyetal04}. 

\section{Splitting of the metric and convenient coordinates}

\label{sec_splitting}

For our purposes, it is convenient to decompose the line element (\ref{ring}) as
\begin{equation}
 \d s^2=\d s_0^2+\lambda\Delta ,
  \label{decomposition}
\end{equation}
in which $\d s^2_0$ is Minkowski spacetime [given by Eq.~(\ref{ring}) with $\lambda=0$, i.e., $F(x)=1=F(y)$], and 
\beqn
 \Delta= \!\! & & \frac{x-y}{1+\lambda x}\d t^2+\frac{L^2}{(x-y)^2}\left[\frac{x+1}{1-\lambda}(y^2-1)\d\psi^2\right. \nonumber \label{perturbation}\\
         & & \quad {}+\left.\frac{x-y}{1+\lambda y}\frac{\d y^2}{y^2-1}+\frac{x+1}{1-\lambda}(1-x^2)\d\phi^2\right] 
\eeqn
measures the deviation from flatness of the full black ring metric~(\ref{ring}). Asymptotically ($x,y\to -1$), $\Delta$ becomes ``negligible'' (in the sense of the Minkowskian metric $\d s_0^2$). 

A boost is now naturally defined with respect to the flat background $\d s_0^2$ (as well as with respect to asymptotic infinity), namely by its isometries. We wish to visualize this in standard cartesian coordinates. In order to introduce them, it is first convenient to replace the ``$C$-metric'' coordinates $(y,x)$ with $(\xi,\eta)$ via the substitution\footnote{We have simply inverted the relations  $\xi=L\sqrt{y^2-1}/(x-y)$ and $\eta=L\sqrt{1-x^2}/(x-y)$ of \cite{EmpRea02prd}.} 
\beqn
 y=-\frac{\xi^2+\eta^2+L^2}{\sqrt{(\xi^2+\eta^2-L^2)^2+4L^2\eta^2}} \, , \nonumber \label{cylindrical} \\
 x=-\frac{\xi^2+\eta^2-L^2}{\sqrt{(\xi^2+\eta^2-L^2)^2+4L^2\eta^2}} \, .
\eeqn
The flat term $\d s_0^2$ in Eq.~(\ref{decomposition}) then takes the form of Minkowski space in double cylindrical coordinates
\be
 \d s_0^2=-\d t^2+\d\eta^2+\eta^2\d\phi^2+\d\xi^2+\xi^2\d\psi^2 ,
 \label{backcyl}
\ee
and the additional quantity $\Delta$ reads
\beqn
  & & \Delta=\frac{2L^2}{\Sigma(1+\lambda x)}\d t^2+2L^2\frac{\left[(\xi^2-\eta^2-L^2)\d\xi+2\eta\xi\d\eta\right]^2}{\Sigma^3(1+\lambda y)} \nonumber \label{pert2} \\
 & & \qquad \qquad {}+\frac{\Sigma-\eta^2-\xi^2+L^2}{\Sigma(1-\lambda)}\left(\xi^2\d\psi^2+\eta^2\d\phi^2\right) ,
\eeqn
where we have denoted
\be
 \Sigma=\sqrt{(\xi^2+\eta^2-L^2)^2+4L^2\eta^2} .
 \label{Sigma}
\ee
In Eq.~(\ref{pert2}) we have kept an explicit simple dependence on the old coordinates $(y,x)$ for brevity and for later convenience [but one can readily substitute Eq.~(\ref{cylindrical}) into Eq.~(\ref{pert2}) if necessary].

Cartesian coordinates are finally given by
\beqn
  & x_1=\eta\cos\phi , \qquad & y_1=\xi\cos\psi , \nonumber \label{cartesian} \\ 
  & x_2=\eta\sin\phi , \qquad & y_2=\xi\sin\psi ,
\eeqn
so that $\eta=\sqrt{x_1^2+x_2^2}$, $\xi=\sqrt{y_1^2+y_2^2}$, and the background is $\d s_0^2=-\d t^2+\d x_1^2+\d x_2^2+\d y_1^2+\d y_2^2$. 

In principle, one could now study a boost along a general direction. Since the original spacetime (\ref{ring}) is symmetric under (separate) rotations in the $(x_1,x_2)$ and $(y_1,y_2)$ planes, such a direction can be specified by a single parameter~$\alpha$, namely introducing the rotated axes 
\be
 z_1=x_1\cos\alpha+y_1\sin\alpha , \quad z_2=-x_1\sin\alpha+y_1\cos\alpha .
 \label{rotated}
\ee
Defining suitable double null coordinates $(u',v')$ by 
\begin{equation}
 t=\frac{-u'+v'}{\sqrt{2}} , \qquad z_1=\frac{u'+v'}{\sqrt{2}} ,
 \label{nullcoords}
\end{equation}
a Lorentz boost along $z_1$ takes the simple form
\begin{equation}
 u'=\epsilon^{-1}u  , \qquad v'=\epsilon v  .
 \label{lorentzboost}
\end{equation}
The parameter $\epsilon>0$ is related to the standard Lorentz factor via $\gamma=(\epsilon+\epsilon^{-1})/2$. 

In the following, we will study in detail two different boosts of the black ring along the privileged axes $x_1$ ($\alpha=0$) and $y_1$ ($\alpha=\pi/2$), which are respectively ``orthogonal'' and ``parallel'' to the ring. But we will also discuss a boost in a general direction. In particular, we will consider ``ultrarelativistic'' boosts to the speed of light, i.e. the transformation~(\ref{lorentzboost}) in the limit $\e\to 0$. Along with that, we will perform the standard mass rescaling~\cite{AicSex71}
\begin{equation}
 M=\gamma^{-1}p=2\epsilon(1+\epsilon^2)^{-1}p ,
 \label{ASmassrescaling}
\end{equation}
which keeps the total energy finite ($p>0$ is a constant). From Eq.~(\ref{mass}), in term of the dimensionless parameter $\lambda$ the rescaling~(\ref{ASmassrescaling}) becomes 
\be
 \lambda=\lambda_\e=\epsilon \frac{8p}{3\pi L^2+\epsilon(8p+3\pi L^2\epsilon)} ,
 \label{ASlambda}
\ee
so that when $\e\to 0$ then $\lambda_\e\approx\e\frac{8p}{3\pi L^2}\to0$.

\section{Orthogonal boost: $\alpha=0$}

\label{sec_orthogonal}

\subsection{Evaluation of the impulsive limit of the metric}

For $\alpha=0$ in Eq.~(\ref{rotated}), Eq.~(\ref{nullcoords}) reduces to
\be
 t=\frac{-u'+v'}{\sqrt{2}} , \qquad x_1=\frac{u'+v'}{\sqrt{2}} ,
 \label{nullx1}
\ee
so that the transformation (\ref{lorentzboost}) describes a Lorentz boost along the $x_1$ axis, which lies in the 2-plane spanned by $(\eta,\phi)$ [cf. Eq.~(\ref{cartesian})]. The latter is orthogonal to the \mbox{2-plane} $(\xi,\psi)$, which contains the ring circle. We wish now to evaluate how the black ring metric~(\ref{ring}) [that is, Eq.~(\ref{decomposition}) with Eqs.~(\ref{backcyl}) and (\ref{pert2})] transforms under the boost~(\ref{lorentzboost}) with $\alpha=0$. Since the coordinates $\xi$ and $\psi$ remain unchanged in this case, it suffices to substitute only the first column of Eq.~(\ref{cartesian}) into Eqs.~(\ref{backcyl}) and (\ref{pert2}). Then, we put Eq.~(\ref{nullx1}) into the thus obtained expressions for $\d s_0^2$ and for $\Delta$ (we omit the intermediate expressions, which are cumbersome and not of particular significance). Finally, we perform the boost~(\ref{lorentzboost}). This leaves $\d s_0^2$ invariant ($2\d u'\d v'=2\d u\d v$), i.e. 
\be
 \d s_0^2=2\d u\d v+\d x_2^2+\d\xi^2+\xi^2\d\psi^2 ,
 \label{back_orth}
\ee  
and makes $\Delta$ dependent parametrically on $\e$. Using the shortcut
\be
 z_\e=\frac{1}{\sqrt{2}}(\e^{-1}u+\e v) ,
 \label{shortcut}
\ee
one obtains\footnote{Again, for convenience in Eq.~(\ref{pert_orth}) we have left some expressions containing the old coordinates $y$ and $x$ [cf. Eq.~(\ref{cylindrical})], which now depend on $\e$. However, these terms will not contribute to the final result in the limit $\e\to 0$.}
\beqn
 \Delta_\e & & \hspace{-.2cm} = \frac{L^2(\e^{-1}\d u-\e\d v)^2}{\Sigma_\e(1+\lambda_\e x)} \nonumber \\ 
  & & {}+\frac{2L^2}{\Sigma_\e^3(1+\lambda_\e y)}\Big(\left[\xi^2-z_\e^2-x_2^2-L^2\right]\d\xi \nonumber \\
  & & \qquad {}+2\xi\left[\textstyle{\frac{1}{\sqrt{2}}}z_\e(\e^{-1}\d u+\e \d v)+x_2\d x_2\right]\Big)^2 \nonumber \label{pert_orth} \\
 & & {}+\frac{\Sigma_\e-z_\e^2-x_2^2-\xi^2+L^2}{\Sigma_\e(1-\lambda_\e)}\Bigg(\xi^2\d\psi^2 \nonumber \\ 
 & & \qquad {}+\frac{\frac{1}{2}\left[\sqrt{2}z_\e\d x_2-x_2(\e^{-1}\d u+\e\d v)\right]^2}{z_\e^2+x_2^2}\Bigg) . 
\eeqn
Here, the quantity $\Sigma_\e$ comes from the expression (\ref{Sigma}) using the above described coordinate transformations and the boost~(\ref{lorentzboost}), and it can be rewritten as
\be
 \Sigma_\e=\sqrt{\left[z_\e^2+x_2^2+(\xi+L)^2\right]\left[z_\e^2+x_2^2+(\xi-L)^2\right]} .
 \label{Sigma_orth}
\ee

We are now interested in taking the ultrarelativistic limit ${\e\to 0}$, i.e. in finding the resulting metric 
\be 
 \d s^2=\d s_0^2+\lim_{\e\to 0}\lambda_\e\Delta_\e .
 \label{limit_orth}
 \ee 
Recalling Eq.~(\ref{ASlambda}), one easily sees that $\lim_{\e\to 0}(\lambda_\e\Delta_\e)=0$ at any given spacetime point with $u \neq 0$ (and away from the ring singularity ${y=-\infty}$). At ${u=0}$ this limit diverges, but in fact it represents a sound distribution supported on ${u=0}$. By inspecting the various quantities in Eq.~(\ref{pert_orth}), it suffices to retain only the terms proportional to $\d u^2$, as the remaining ones become negligible for ${\e\to 0}$. Similarly, we drop the factors $1+\lambda_\e x$, $1+\lambda_\e y$ and $1-\lambda_\e$, since $\lambda_\e\to 0$ for $\e\to 0$. Using Eqs.~(\ref{ASlambda}), (\ref{pert_orth}) and (\ref{shortcut}), for $\e\sim 0$ we can thus write 
\be
 \lambda_\e\Delta_\e\approx\frac{8p}{3\pi L^2}\frac{1}{\e}h_{_{\!\bot}}\!(z_\e)\d u^2 , 
 \label{dominant_orth}
\ee
where 
\beqn
 h_{_{\!\bot}}\!(z_\e)= & & \frac{2L^2-x_2^2}{2\Sigma_\e}+\frac{4L^2\xi^2z^2_\e }{\Sigma^3_\e} +\frac{x_2^2(L^2-\xi^2)}{2(z^2_\e +x_2^2)\Sigma_\e} \nonumber \label{h_orth} \\
  & & \qquad {}+\frac{x_2^2}{2(z^2_\e +x_2^2)} .
\eeqn
We have emphasized the dependence of $h_{_{\!\bot}}$ on $z_\e$ (which gives the only dependence on $\e$) because this is essential in our limit (of course, $h_{_{\!\bot}}$ depends on the coordinates $x_2$ and $\xi$ as well). In taking the limit $\e\to 0$ of Eq.~(\ref{dominant_orth}), we can now apply the distributional identity [recall Eq.~(\ref{shortcut})]
\be
 \lim_{\e\to 0} \frac{1}{\e}f\left(z_\e \right)     
  =\sqrt{2}\,\delta(u)\int_{-\infty}^{+\infty}f(z)\d z .
 \label{identity} 
\ee
With this, the final metric is [cf.~Eqs.~(\ref{limit_orth}) and (\ref{back_orth})]
\be
 \d s^2=2\d u\d v+\d x_2^2+\d\xi^2+\xi^2\d\psi^2+H_{_{\!\bot}}\!(x_2,\xi)\delta(u)\d u^2 ,  
 \label{pporth}
\ee
with a profile function given by 
\be
 H_{_{\!\bot}}\!(x_2,\xi)=\frac{8\sqrt{2}p}{3\pi L^2}\left(\int_{-\infty}^{+\infty}h_{_{\!\bot}}\!(z)\d z\right) .
 \label{integ_orth}
\ee 
It only remains to explicitly perform the integration in Eq.~(\ref{integ_orth}), with $h_{_{\!\bot}}$ given by Eq.~(\ref{h_orth}) with Eq.~(\ref{Sigma_orth}). The last term in Eq.~(\ref{h_orth}) gives rise to the simple integral $\int_{-\infty}^{+\infty}(z^2+x_2^2)^{-1}\d z=\pi|x_2|^{-1}$. The first three terms lead to the elliptic integrals~(\ref{orth1}), (\ref{orth2}) and (\ref{orth3}) of Appendix~\ref{app_integrals}. Combining the various quantities, we finally obtain
\beqn
  & & \hspace{-.7cm} H_{_{\!\bot}}\!(x_2,\xi)=\frac{8\sqrt{2}p}{3\pi L^2} \nonumber \\
 & {}\times & \Bigg[\left(3L^2+\xi^2+x_2^2\frac{\xi+L}{\xi-L}\right)\frac{K(k)}{\sqrt{(\xi+L)^2+x_2^2}} \nonumber \label{H0orth} \\ 
 & & \hspace{-.2cm} {}-\sqrt{(\xi+L)^2+x_2^2}\,E(k) \nonumber \\
 & & \hspace{-.2cm} {}-\frac{\xi+L}{\xi-L}\frac{(\xi-L)^2+x_2^2}{\sqrt{(\xi+L)^2+x_2^2}}\,\Pi(\rho_0,k)+\frac{\pi}{2}|x_2| \Bigg] , 
\eeqn
with
\be
 k=\sqrt{\frac{4\xi L}{(\xi+L)^2+x_2^2}} , \qquad \rho_0=-\frac{(\xi-L)^2}{x_2^2} .
 \label{krho0_orth}
\ee
One can reexpress the elliptic integral $\Pi(\rho_0,k)$ using identities~(\ref{iden1}) and (\ref{iden2}), and obtain an alternative form of $H_{_{\!\bot}}$, which will be useful for subsequent discussions, 
\beqn
 & & \hspace{-.6cm} H_{_{\!\bot}}\!(x_2,\xi)=\frac{8\sqrt{2}p}{3\pi L^2} \nonumber \\
 & {}\times & \Bigg[\frac{3L^2+\xi^2}{\sqrt{(\xi+L)^2+x_2^2}}\,K(k)-\sqrt{(\xi+L)^2+x_2^2}\,E(k) \nonumber \label{Horth}\\ 
 & {}+ & \frac{\xi-L}{\xi+L}\frac{x_2^2}{\sqrt{(\xi+L)^2+x_2^2}}\,\Pi(\rho,k)+\pi|x_2|\Theta(L-\xi) \Bigg] , \nonumber \\
\quad 
\eeqn
where
\be
 \rho=\frac{4\xi L}{(\xi+L)^2} ,
 \label{krho_orth}
\ee
and $\Theta(L-\xi)$ denotes the step function.

Let us observe that no singular coordinate transformation of the type of \cite{AicSex71} had to be performed in the calculation above, since all the required integrals are convergent.

\subsection{Properties of the solution}

\label{subsec_propeties_orth}

\begin{figure}
 \begin{center}
  \includegraphics{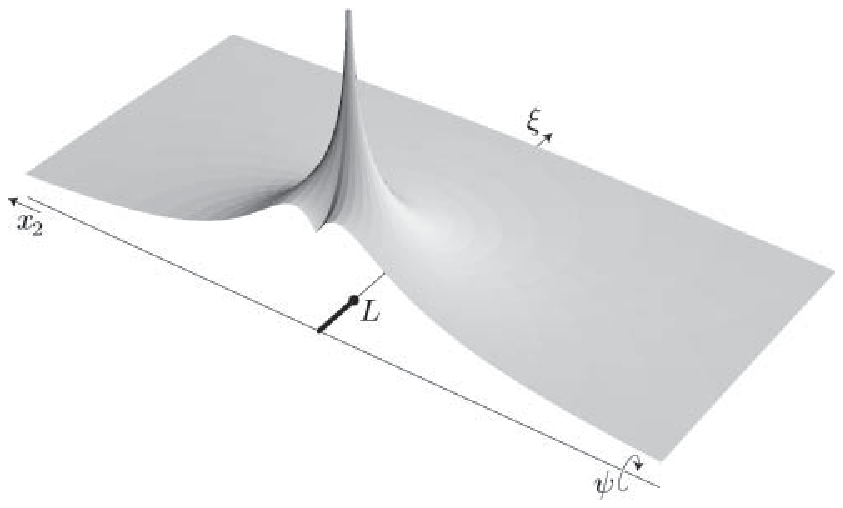}\\[12pt]
  \includegraphics{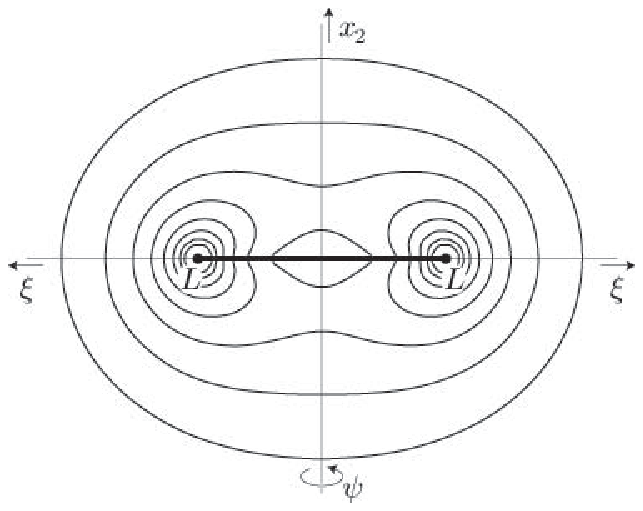}\hspace*{20pt}
 \end{center}
 \caption{The profile function $H_{_{\!\bot}}(x_2,\xi)$ given by Eq.~(\ref{Horth}). The upper picture represents the values taken by $H_{_{\!\bot}}$ over the plane $(x_2,\xi)$, whereas the lower one displays curves along which $H_{_{\!\bot}}$ is constant. This is the case of a static black ring boosted to the speed of light in a direction orthogonal to the plane $(\xi,\psi)$, which contains the ring circle. The coordinates $(x_2,\xi,\psi)$ span spatial sections of the impulsive wave front $u=0$, cf.~Eq.~(\ref{pporth}). Here the coordinate $\psi$ is suppressed, since it just describes the orbits of a Killing vector field. The profile function $H_{_{\!\bot}}$ diverges (only) at the {\em ring} singularity ${x_2=0}$, ${\xi=L}$, as indicated by the thick point(s) in the pictures. In addition, there is a disk membrane within the ring, i.e. at $x_2=0$, $\xi<L$ [cf.~Eq.~(\ref{Ricci_orth})], which manifests itself as a jump in $\pa H_{_{\!\bot}}/\pa x_2$. This is drawn above as a thick line.} 
 \label{fig_orth}
\end{figure}

A static black ring boosted to the speed of light in a direction orthogonal to the ring circle is thus described by the metric~(\ref{pporth}), with $H_{_{\!\bot}}$ given explicitly by Eq.~(\ref{Horth}). This is evidently a five-dimensional impulsive \pp wave propagating along the $x_1$ direction [see Eq.~(\ref{nullx1})]. Such a spacetime is flat everywhere except on the null hyperplane $u=0$, which represents the impulsive wave front. In particular, the line element~(\ref{pporth}) is singular at the points satisfying $u=0=x_2$ and $\xi=L$ [$k=1$ in Eq.~(\ref{krho0_orth})], i.e. on a {\em circle of radius $L$} contained within the wave front. This is a remnant of the curvature singularity ($y=-\infty$) of the original static black ring (\ref{ring}). Since the boost performed above was orthogonal to the ring circle, the latter has not Lorentz-contracted. We have plotted the profile function $H_{_{\!\bot}}$ in Fig.~\ref{fig_orth}. 

\subsubsection{Killing vectors}

The \pp wave line element~(\ref{pporth}), (\ref{Horth}) is obviously invariant under the transformations generated by the vector fields $\pa_v$ and  $\pa_\psi$. It has been demonstrated in four spacetime dimensions \cite{AicBal96} that impulsive \pp waves admit more isometries than the same class of waves with a general profile. Similarly, it is easy to see that, thanks to the presence of $\delta(u)$, the line element~(\ref{pporth}) admits also the three commuting Killing vectors
\be
 u\pa_{x_2}-x_2\pa_v , \quad u\pa_{y_1}-y_1\pa_ v , \quad u\pa_{y_2}-y_2\pa_v . 
 \label{killingorth}
\ee
[Recall the simple relation (\ref{cartesian}) between $(y_1,y_2)$ and $(\xi,\psi)$.] These are generators of null rotations. Incidentally, we observe that impulsive waves in the four-dimensional (anti-)de~Sitter universe can be described as a submanifold of  five-dimensional impulsive \pp waves, and they admit symmetries very similar to the above \cite{PodOrt01}. 

\subsubsection{The Ricci tensor}

The static ring (\ref{ring}) is a vacuum spacetime ($R_{\mu\nu}=0$) everywhere except on the disk membrane $x=+1$ (and of course on the ring singularity $y=-\infty$). Therefore, one would expect also the ultrarelativistic boosted ring to be a vacuum solution except at a possible remnant (after the boost) of the original disk membrane. To check the results, we have verified that the Ricci tensor associated to the spacetime~(\ref{pporth}), (\ref{Horth}) is indeed zero everywhere but at $u=0=x_2$, $\xi<L$, i.e. inside a two-dimensional disk lying on the wave front. Namely, using Eqs.~(\ref{differential}) one finds
\be
 R_{uu}=-\frac{1}{2}{\bf\Delta}H_{_{\!\bot}}\delta(u)=-\frac{8\sqrt{2}p}{3L^2}\,\Theta(L-\xi)\delta(x_2)\delta(u) 
 \label{Ricci_orth}
\ee
[the symbol ${\bf\Delta}$ denotes the Laplace operator over the transverse flat space $(x_2,\xi,\psi)$]. 
This nonvanishing component arises only due to the last term in Eq.~(\ref{Horth}), a typical term associated to boosted conical singularities \cite{LouSan91}. 
On the disk rim $u=0=x_2$, $\xi=L$ the metric~(\ref{pporth}), (\ref{Horth}) is singular, and its exact structure may be not reflected correctly by Eq.~(\ref{Ricci_orth}).

\subsubsection{The Weyl tensor}

For any five-dimensional \pp wave written in the form ${\d s^2=2\d u\d v+\d x_2^2+\d\xi^2+\xi^2\d\psi^2+H_{_{\!\bot}}\d u^2}$, in the null/orthonormal frame  
\beqn
 & & \mbox{\boldmath$k$}=\d u, \qquad \mbox{\boldmath$l$}=-\d v-\frac{1}{2}H_{_{\!\bot}}\d u, \nonumber \label{frame_orth} \\
 & & \mbox{\boldmath$m$}_{(1)}=\d\xi, \quad \mbox{\boldmath$m$}_{(2)}=\xi\d\psi, \quad \mbox{\boldmath$m$}_{(3)}=\d x_2 , 
\eeqn
the Weyl tensor is
\beqn
\mbox{\boldmath$C$}=\Psi_{ij}[(\mbox{\boldmath$k$}\wedge\mbox{\boldmath$m$}_{(i)}) \!& \otimes & \! (\mbox{\boldmath$k$}
  \wedge\mbox{\boldmath$m$}_{(j)}) \nonumber \label{Weyl} \\ 
  {}+(\mbox{\boldmath$k$}\wedge\mbox{\boldmath$m$}_{(j)}) \!& \otimes & \! (\mbox{\boldmath$k$}
  \wedge\mbox{\boldmath$m$}_{(i)})] ,
\eeqn
where summation over $i,j=1,2,3$ is understood. This is the canonical form of type N spacetimes \cite{Coleyetal04,Pravdaetal04}, and $\mbox{\boldmath$k$}$ is the unique principal null direction. The symbols \cite{Pravdaetal04} 
\be
 \Psi_{ij}=\frac{1}{2}C_{\mu\nu\rho\sigma}l^\mu m_{(i)}^\nu l^\rho m_{(j)}^\sigma , \qquad i,j=1,2,3
\ee
define a $3\times 3$ symmetric traceless matrix that expresses the independent frame components of the Weyl tensor, which are in general five in $D=5$.\footnote{The quantities $\Psi_{ij}$ can be understood as a generalization of the complex scalar $\Psi_4$, which fully characterizes type N spacetimes in the well known $D=4$ theory.} In particular, this demonstrates that in the ultrarelativistic boost studied above the original type I$_i$ \cite{PraPra05} of the static ring (\ref{ring}) has degenerated to the type N on the wave front of our specific \pp wave~(\ref{pporth}), (\ref{Horth}). Moreover, for such a solution the symmetry under $\pa_\psi$ implies $\Psi_{12}=0=\Psi_{23}$. One is thus left with
\beqn
 \Psi_{11} & = & -\frac{1}{2}\left(\frac{1}{2}\frac{\pa^2H_{_{\!\bot}}}{\pa\xi^2}\delta(u)+\frac{1}{3}R_{uu}\right) , \nonumber \\ 
 \Psi_{13} & = & -\frac{1}{4}\frac{\pa^2H_{_{\!\bot}}}{\pa\xi\pa x_2}\delta(u) , \nonumber \label{Weylscal} \\
 \Psi_{22} & = & -\frac{1}{2}\left(\frac{1}{2\xi}\frac{\pa H_{_{\!\bot}}}{\pa\xi}\delta(u)+\frac{1}{3}R_{uu}\right) , \\
 \Psi_{33} & = & -\frac{1}{2}\left(\frac{1}{2}\frac{\pa^2H_{_{\!\bot}}}{\pa x_2^2}\delta(u)+\frac{1}{3}R_{uu}\right) . \nonumber 
\eeqn
The above components of the Weyl tensor confirm the presence of an impulsive gravitational wave at $u=0$. For $H_{_{\!\bot}}$ given by Eq.~(\ref{Horth}), the explicit form of the scalars $\Psi_{ij}$ is presented in Appendix~\ref{app_weyl}. There one can observe that the elliptic integral $\Pi(\rho,k)$ disappears from such expressions. 

\subsubsection{Asymptotic behaviour}

The spacetime~(\ref{pporth}) is flat everywhere except on the wave front $u=0$. If we restrict within the latter, it is interesting to analyze how the gravitational field generated by the boosted black ring behaves {\em at a large spatial distance from the centre of the ring singularity} (given by $\xi=0=x_2$). Spatial sections of the wave front are three-dimensional spaces, in which we can introduce standard spherical coordinates $(r,\theta,\psi)$ by
\be
 x_2=r\cos\theta , \qquad \xi=r\sin\theta .
\ee
Since $r$ is a radial coordinate from the centre of the ring ($r^2=x_2^2+\xi^2$) and $L$ is the radius of the ring, we consider an expansion for small values of the dimensionless parameter $L/r$. This means considering Eq.~(\ref{Horth}) for $k$ and $\rho$ approaching zero. Using Eqs.~(\ref{serK1}), (\ref{serE1}) and (\ref{serPi1}), we obtain
\be
 H_{_{\!\bot}}=\frac{1}{\sqrt{2}}\frac{8p}{3L}\left[3\frac{L}{r}-\frac{7}{8}(3\cos^2\theta-1)\frac{L^3}{r^3}+
      O\left(\frac{L^5}{r^5}\right)\right] .
 \label{multipole_orht}
\ee
We recognize the standard multipole terms [indeed, for $\xi>L$, $H_{_{\!\bot}}$ is a solution of a three-dimensional Laplace equation, cf.~Eq.~(\ref{Ricci_orth})]. Notice that the dipole term is missing, due to the geometry of the source. In the limit when the ring shrinks to a point, i.e. $L\to 0$, the expansion reduces just to the monopole term,
\be
 H_{_{\!\bot}}^0=\lim_{L\to 0} H_{_{\!\bot}}=\frac{1}{\sqrt{2}}\frac{8p}{r} .
 \label{monopole}
\ee
This exhibits the ``Newtonian'' $1/r$ fall-off in three-dimensional space, with a ``mass'' proportional to $p$. The metric~(\ref{pporth}) with a profile function given by $H_{_{\!\bot}}^0$ coincides with the five-dimensional analogue of the \AS solution, obtained by boosting the Schwarzschild line element to the speed of light \cite{LouSan90} (cf. also, e.g., \cite{Ortaggio05,EarGid02,YosNam02}).

In order to gain further physical insight, one can similarly consider other expansions near ``special places''. For example, {\em near the axis $\xi=0$} we obtain 
\beqn
 H_{_{\!\bot}}= & & \frac{8\sqrt{2}p}{3L^2}\left[\frac{L^2-x_2^2}{\sqrt{L^2+x_2^2}}+|x_2|\right. \nonumber \\
 & & \quad {}+\left.\frac{3L^2}{4}\frac{L^2-x_2^2}{\sqrt{(L^2+x_2^2)^5}}\,\xi^2+O\left(\xi^4\right)\right] .
\eeqn

{\em Near the plane of the ring $x_2=0$},
\beqn
 H_{_{\!\bot}}= & & \frac{1}{\sqrt{2}}\frac{8p}{3\pi L^2}\left[2\frac{3L^2+\xi^2}{L+\xi}K(\tilde k)-2(L+\xi)E(\tilde k)\right. \nonumber \\
 & & {}+2\pi|x_2|\Theta(L-\xi)-\left(\frac{1}{L+\xi}K(\tilde k)\right. \nonumber \\ 
 & & {}+\left.\left.\frac{5L^2-\xi^2}{(L-\xi)^2(L+\xi)}E(\tilde k)\right)x_2^2+O\left(x_2^4\right)\right] ,
\eeqn
where $\tilde k=k(x_2=0)=\sqrt{4L\xi}/(L+\xi)$.

If we introduce suitable coordinates ``centred on the ring''
\be
 x_2=\tilde r\sin\tilde\theta , \qquad \xi=L+\tilde r\cos\tilde\theta ,
\ee
using Eqs.~(\ref{serK2}), (\ref{serE2}) and (\ref{serPi2}), the expansion [of Eq.~(\ref{H0orth})] {\em near the singular ring $\tilde r=0$} is
\beqn
 \hspace{-.6cm} H_{_{\!\bot}}= & & \frac{1}{\sqrt{2}}\frac{8p}{3\pi L}\left\{-4\left(1+\log\frac{\tilde r}{8L}\right)+2\tilde\theta\sin\tilde\theta\frac{\tilde r}{L}\right. \nonumber \\
 & & \hspace{-1cm} {}-\left.\frac{1}{4}\left[1+\cos2\tilde\theta\left(1+3\log\frac{\tilde r}{8L}\right)\right]\frac{\tilde r^2}{L^2}+
      O\left(\frac{\tilde r^3}{L^3}\right)\right\} .
\eeqn
The ring-shaped singularity is explicitly visible in the first logarithmic term.

\section{Parallel boost: $\alpha=\pi/2$}

\label{sec_parallel}

\subsection{Evaluation of the impulsive limit of the metric}

\label{subsec_evalimpulslimit}

For $\alpha=\pi/2$ in Eq.~(\ref{rotated}), Eq.~(\ref{nullcoords}) becomes 
\be
 t=\frac{-u'+v'}{\sqrt{2}} , \qquad y_1=\frac{u'+v'}{\sqrt{2}} ,
 \label{nully1}
\ee
so that the transformation (\ref{lorentzboost}) describes a boost in the $y_1$ direction, i.e. in the 2-plane $(\xi,\psi)$ [cf. Eq.~(\ref{cartesian})] containing the ring circle. As for the previous orthogonal boost, we need to calculate how the black ring metric~(\ref{ring}) transforms under the boost~(\ref{lorentzboost}), and then take the limit $\e\to 0$. In the present case ($\alpha=\pi/2$) the coordinates $\eta$ and $\phi$ remain unchanged, hence we substitute the second column of Eq.~(\ref{cartesian}) into Eqs.~(\ref{backcyl}) and (\ref{pert2}). Apart from this, we follow the same steps as in Sec.~\ref{sec_orthogonal}, and the derivation here will be therefore shortened in its straightforward parts. The flat, boost-invariant part of the decomposition~(\ref{decomposition}) can now be written as 
\be 
 \d s_0^2=2\d u\d v+\d y_2^2+\d\eta^2+\eta^2\d\phi^2 .
\ee  
For the additional term $\lambda\Delta$, as ${\e\sim 0}$ we get an expression analogous to Eq.~(\ref{dominant_orth}), but with $h_{_{\!\bot}}\!(z_\e)$ replaced by 
\beqn
 h_{_{||}}\!(z_\e)= & & \frac{4L^2-y_2^2}{2\Sigma_\e }-\frac{4L^2\eta^2z_\e^2}{\Sigma^3_\e }-
  \frac{y_2^2(L^2+\eta^2)}{2(z_\e^2 +y_2^2)\Sigma_\e } \nonumber 
  \label{h_parall} \\
  & & {}+\frac{y_2^2}{2(z^2_\e+y_2^2)} ,
\eeqn
and
\be
 \Sigma_\e=\sqrt{z_\e^4+2(y_2^2+\eta^2-L^2)z_\e^2+a^4} , \label{Sigma_parall}
\ee
with
\be
 a=\left[(\eta^2+y_2^2-L^2)^2+4\eta^2L^2\right]^{1/4} \label{def_a}.
\ee
Again employing identity~(\ref{identity}), the final boosted metric is now
\be
 \d s^2=2\d u\d v+\d y_2^2+\d\eta^2+\eta^2\d\phi^2+H_{_{\!||}}\!(y_2,\eta)\delta(u)\d u^2 ,  
 \label{ppparall}
\ee
with a profile function
\be
 H_{_{\!||}}\!(y_2,\eta)=\frac{8\sqrt{2}p}{3\pi L^2}\left(\int_{-\infty}^{+\infty}h_{_{||}}\!(z)\d z\right) .
 \label{integ_parall}
\ee 
We employ the elliptic integrals~(\ref{parall1}), (\ref{parall2}) and (\ref{parall3}) of Appendix~\ref{app_integrals} in order to perform the integration in Eq.~(\ref{integ_parall}) [with $h_{_{||}}$ given by Eqs.~(\ref{h_parall}) and (\ref{Sigma_parall})].  Combining all the terms, and using identity~(\ref{iden2}) to reexpress the elliptic integral $\Pi$ in a more convenient form, we obtain
\beqn
 & & \hspace{-.7cm} H_{_{\!||}}\!(y_2,\eta)=\frac{8\sqrt{2}p}{3\pi L^2}\Bigg[\frac{5L^2+\eta^2+2a^2}{2a}K(k)-2aE(k) \nonumber \label{Hparall} \\ 
 & & \hspace{-.2cm}     
      {}+\frac{\eta^2+L^2}{2a}\frac{a^2+y_2^2}{a^2-y_2^2}\,\Pi(\rho,k)+\pi|y_2|\Theta\left(y_2^2-a^2\right)\Bigg] , 
\eeqn
where
\beqn
 k & = & \frac{\left(a^2-\eta^2-y_2^2+L^2\right)^{1/2}}{\sqrt{2}a} , \nonumber \label{krho_parall} \\
\rho & = & -2y_2^2\frac{a^2-\eta^2-y_2^2+L^2}{\left(a^2-y_2^2\right)^2} ,
\eeqn
and $a$ as in Eq.~(\ref{def_a}).

\subsection{Properties of the solution}

\label{subsec_propeties_parall}

\begin{figure}
 \begin{center}
  \includegraphics{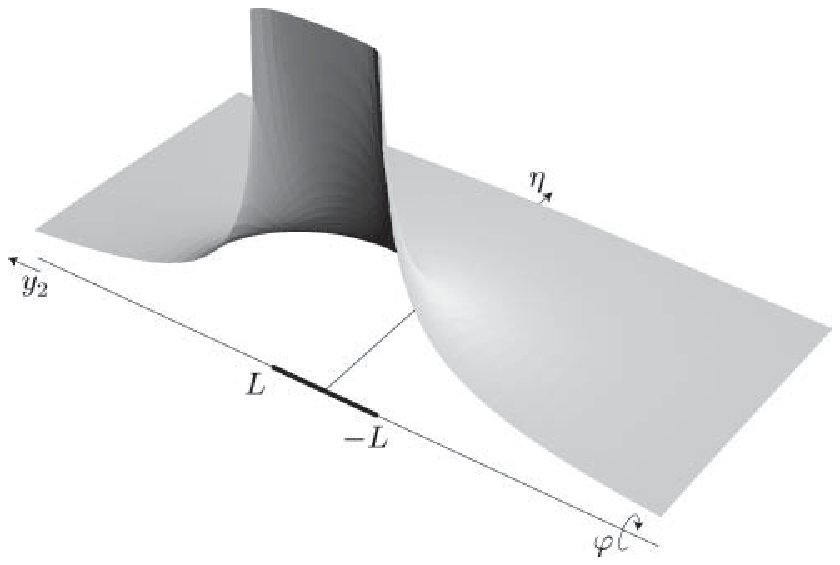}\\[12pt]
  \includegraphics{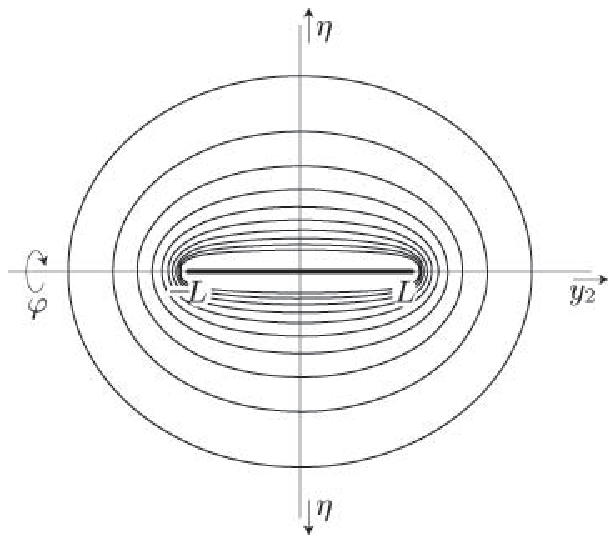}\hspace*{20pt}
 \end{center}
 \caption{Plot of the profile function $H_{_{\!||}}(y_2,\eta)$ given by Eq.~(\ref{Hparall}).  This is the case of a static black ring boosted to the speed of light along a direction contained within the plane of the ring circle, i.e. $(\xi,\psi)$. The coordinates $(y_2,\eta,\phi)$ span spatial sections of the impulsive wave front $u=0$, cf.~Eq.~(\ref{ppparall}). The Killing coordinate $\phi$ is suppressed in the figures. The profile function $H_{_{\!||}}$ diverges (only) at the {\em rod} singularity $\eta=0$, $|y_2|\le L$, as indicated by the thick line in the pictures.} 
 \label{fig_parall}
\end{figure}

A static black ring boosted to the speed of light in a direction contained in the plane of the ring circle is thus represented by the metric~(\ref{ppparall}), (\ref{Hparall}). As in the case of the orthogonal boost of Sec.~\ref{sec_orthogonal}, this is a five-dimensional impulsive \pp wave. It propagates along the $y_1$ direction, and it is singular at the points satisfying $u=0=\eta$ and $|y_2|\le L$ [$k=1$ in Eq.~(\ref{krho_parall})], i.e. on a {\em rod of length $2L$} contained within the wave front. This is a remnant of the curvature singularity of the original static black ring (\ref{ring}), which has Lorentz-contracted because of the ultrarelativistic boost in the plane of the ring. On the contrary, notice that the apparent divergence of $H_{_{\!||}}$ at $y_2^2=a^2$ is only a fictitious effect: the singular behaviour of the coefficient of $\Pi$ in Eq.~(\ref{Hparall}) is compensated if one takes into account the form of $\rho$ [Eq.~(\ref{krho_parall})] and the step function in the last term. The profile function $H_{_{\!||}}$ is plotted in Fig.~\ref{fig_parall}. 

The discussion of further properties of the solution~(\ref{ppparall}), (\ref{Hparall}) is now shortened, since it follows the similar one in Sec.~\ref{sec_orthogonal}. There exist isometries generated by the Killing vector fields $\pa_v$, $\pa_\phi$, $u\pa_{y_2}-y_2\pa_v$, $u\pa_{x_1}-x_1\pa_v$ and $u\pa_{x_2}-x_2\pa_v$ [cf. Eq.~(\ref{killingorth})].

During the parallel boost, also the original disk membrane has Lorentz-contracted, and it is now located on the singular region $u=0=\eta$, $|y_2|\le L$. We will not discuss the behaviour of the solution there. Except on this singular rod, the Ricci tensor associated to the spacetime~(\ref{ppparall}), (\ref{Hparall}) is vanishing, as we verified using identities~(\ref{differential}). 

Similarly as in Sec.~\ref{sec_orthogonal}, one can cast the Weyl tensor in the type N canonical form using 
the frame~(\ref{frame_orth}) with the replacements $\xi\to\eta$, $\psi\to\phi$, $x_2\to y_2$ and, of course, $H_{_{\!\bot}}\to H_{_{\!||}}$. Analogously, one obtains the corresponding Weyl components from Eq.~(\ref{Weylscal}). In this case, we omit the explicit form of the scalars $\Psi_{ij}$, which is rather complicated and does not provide any immediate physical insight. We just notice that, again, the elliptic integral $\Pi$ disappears. 

Following the corresponding analysis of Sec.~\ref{sec_orthogonal}, we can analyze how the gravitational field generated by the boosted black ring~(\ref{ppparall}) behaves {\em at a large spatial distance from the centre of the rod singularity} (given by ${\eta=0=y_2}$). With spherical coordinates defined on the wave front by
\be
 y_2=r\cos\theta , \qquad \eta=r\sin\theta ,
\ee
we now obtain 
\be
 H_{_{||}}=\frac{1}{\sqrt{2}}\frac{8p}{3L}\left[3\frac{L}{r}+\frac{5}{8}(3\cos^2\theta-1)\frac{L^3}{r^3}+
      O\left(\frac{L^5}{r^5}\right)\right] .
\ee
The monopole term coincides with that obtained in the case of $H_{_{\bot}}$, cf. Eqs.~(\ref{multipole_orht}) and (\ref{monopole}), and for $L\to 0$ (i.e., when the rod shrinks to a point) it gives rise to the five-dimensional \AS solution. Again, there is no dipole, but the quadrupole term is different from that of Eq.~(\ref{multipole_orht}).

In addition, we can consider an expansion of $H_{_{||}}$ {\em near the axis $\eta=0$}. The rod singularity lies exactly at $\eta=0$, for $|y_2|\le L$. Therefore, we have to study the two cases $|y_2|>L$ and $|y_2|<L$ separately. For $|y_2|>L$, one has 
\beqn
 H_{_{||}}= & & \frac{8\sqrt{2}p}{3L^2}\left[\frac{L^2}{\sqrt{y_2^2-L^2}}+|y_2|-\sqrt{y_2^2-L^2}\right. \nonumber \\
 & & \qquad {}-\left.\frac{3}{4}\frac{L^2y_2^2}{\sqrt{(y_2^2-L^2)^5}}\,\eta^2+O\left(\eta^4\right)\right] .
\eeqn 
The case $|y_2|<L$ is more delicate and one has to employ expansions~(\ref{serK2}), (\ref{serE2}) and (\ref{serPi2}). At the end, 
\beqn
 H_{_{||}}= & & \frac{8\sqrt{2}p}{3\pi L^2}\Bigg[\frac{y_2^2-2L^2}{\sqrt{L^2-y_2^2}}\log\frac{L^2\eta^2}{16(L^2-y_2^2)^2}-2\sqrt{L^2-y_2^2} \nonumber \\
 & & \quad {}+|y_2|\mbox{arccot}\frac{L^2-2y_2^2}{2|y_2|\sqrt{L^2-y_2^2}}+O\left(\eta^2\right)\Bigg] ,
\eeqn
where ``arccot'' takes values in $[0,\pi]$. The first term carries the singular behaviour at the rod $\eta=0$. 

\begin{figure*}
 \begin{center}
 \includegraphics{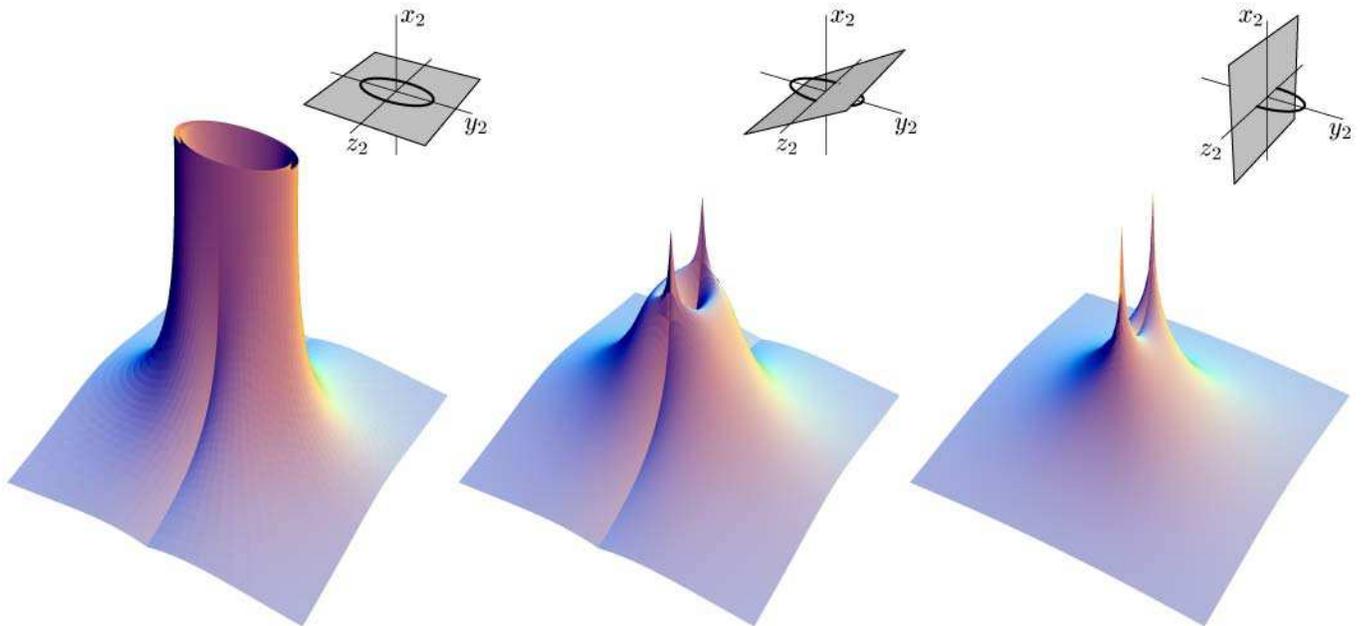}\hspace*{20pt}
 \end{center}
 \caption{Plot of the profile function $H(x_2,y_2,z_2)$ given by Eqs.~(\ref{integ_general}) and (\ref{h_general}). The integration in Eq.~(\ref{integ_general}) has been performed numerically. The coordinates $(x_2,y_2,z_2)$ span spatial sections of the impulsive wave front $u=0$, cf.~Eq.~(\ref{ppgeneral}). Since this plot corresponds to a boost in a general direction $z_1$, the function $H$ is not axially symmetry. We have thus depicted representative plots of the values taken by $H$ over different sections of the three space $(x_2,y_2,z_2)$. In the left figure, in particular, it is evident the ellipse-shaped singularity, cf. Eq.~(\ref{ellipse}).} 
 \label{fig_general}
\end{figure*}

\section{General boost: an arbitrary $\alpha$}

\label{sec_general}

We finally consider the boost in a general direction~$z_1$, which is characterized by the
angular parameter~$\alpha$, see Eqs.~(\ref{rotated})--(\ref{lorentzboost}). We employ the method of the previous sections, and after straightforward calculations we again obtain an impulsive \pp wave  
\be
 \d s^2=2\d u\d v+\d x_2^2+\d y_2^2+\d z_2^2+H(x_2,y_2,z_2)\delta(u)\d u^2 .  
 \label{ppgeneral}
\ee
Now the profile function
\be
 H(x_2,y_2,z_2)=\frac{8\sqrt{2}p}{3\pi L^2}\left(\int_{-\infty}^{+\infty}h(z)\d z\right) ,
 \label{integ_general}
\ee 
is an integral of the function
\beqn
 h(z)= & & \frac{L^2}{\Sigma}+\frac{L^2}{\Sigma^3}\left[(\xi^2-\eta^2-L^2)\frac{y_{1}}{\xi}\sin\alpha+2\xi x_{1}\cos\alpha\right]^2\nonumber \label{h_general} \\
  & & \hspace{-10mm}{}+\frac{1}{2}\left(1-\frac{\xi^2+\eta^2-L^2}{\Sigma}\right)
\left(\frac{y_2^2}{\xi^2}\sin^2\alpha+\frac{x_2^2}{\eta^2}\cos\alpha\right) .  
\eeqn
Here the dependence on $z$ is contained in
\beqn
 & & y_1=z\sin\alpha+z_2\cos\alpha, \qquad \xi^2=y_1^2+y_2^2 , \nonumber \label{xieta} \\
 & & x_1=z\cos\alpha-z_2\sin\alpha, \qquad \eta^2=x_1^2+x_2^2 ,
\eeqn
and in $\Sigma$, given by Eq.~(\ref{Sigma}). In order to perform the above integration, it is convenient to factorize $\Sigma$ as
\be
 \Sigma=\sqrt{\left[(z-r_1)^2+s_1^2\right]\left[(z+r_1)^2+s_2^2\right]} , \label{Sigma_general decomp}
\ee
%where the parameters $r_1$, $s_1$ and $s_2$ are defined by
%\beqn
% & & s_1^2=r_1^2+A-\frac{2B}{r_1} , \quad s_2^2=r_1^2+A+\frac{2B}{r_1} , \nonumber \label{system} \\
%% s_2^2 &=& r_1^2+A+\frac{2B}{r_1} , \label{system} \\
% & &  r_1^6+Ar_1^4+{\textstyle\frac{1}{4}}(A^2-C)r_1^2-B^2=0 , 
%\eeqn
%with
%\beqn
% A &=& x_2^2+y_2^2+z_2^2-L^2+2L^2\cos^2\alpha , \nonumber \\
% B &=& L^2z_2\sin\alpha\cos\alpha , \label{ABC} \\
% C &=& (x_2^2+y_2^2+z_2^2-L^2)^2+4L^2(x_2^2+z_2^2\sin^2\alpha) . \nonumber
%\eeqn
where the parameters $s_1$ and $s_2$ are defined by
\be
 s_1^2=r_1^2+A-\frac{2B}{r_1} , \qquad s_2^2=r_1^2+A+\frac{2B}{r_1} , \label{system}
\ee
and $r_1$ by the equation (of third order in $r_1^2$) 
\be
 r_1^6+Ar_1^4+{\textstyle\frac{1}{4}}(A^2-C)r_1^2-B^2=0 ,
\ee
with
\beqn
 A &=& x_2^2+y_2^2+z_2^2-L^2+2L^2\cos^2\alpha , \nonumber \\
 B &=& L^2z_2\sin\alpha\cos\alpha , \label{ABC} \\
 C &=& (x_2^2+y_2^2+z_2^2-L^2)^2+4L^2(x_2^2+z_2^2\sin^2\alpha) . \nonumber
\eeqn
Using Cardano's formula we may write the root $r_1^2$ as 
\be
 r_1^2=-\frac{A}{3}+\sqrt[3]{-\frac{q}{2}+\sqrt{Q}} + \sqrt[3]{-\frac{q}{2}-\sqrt{Q}} , \label{Cardano}
\ee
where
\beqn
 Q &=& \left(\frac{p}{3}\right)^3+\left(\frac{q}{2}\right)^2 , \nonumber\\
 p &=& -{\textstyle\frac{1}{3}}A^2+{\textstyle\frac{1}{4}}(A^2-C) , \label{Qpq}\\
 q &=& {\textstyle\frac{2}{27}}A^3-{\textstyle\frac{1}{12}}A(A^2-C)-B^2. \nonumber
\eeqn
Notice that for the particular case of the orthogonal boost ($\alpha=0$) we obtain 
$r_1=0$, $s_1^2=x_2^2+(\xi+L)^2$, $s_2^2=x_2^2+(\xi-L)^2$, which coincides with Eq.~(\ref{Sigma_orth}), while
for the parallel boost ($\alpha=\frac{\pi}{2}$) one has $2r_1^2=a^2-y_2^2-\eta^2+L^2$ and  $2s_1^2=2s_2^2=a^2+y_2^2+\eta^2-L^2$, which is equivalent to Eq.~(\ref{Sigma_parall}) [$a$ is defined in Eq.~(\ref{def_a})]. For any $\alpha$, the integral~(\ref{integ_general}) could now be expressed using elliptic integrals, in principle (because $\Sigma$ is a square root of a fourth order polynomial in $z$) \cite{Gradshteynbook6,GroHofbook2}. For example, the simplest first term in Eq.~(\ref{h_general}) leads to \cite{GroHofbook2}
\be
\int_{-\infty}^{+\infty}\frac{\d z}{\Sigma(z)} =\frac{2}{\sqrt{s_1s_2k_1}}\,K(k),
 \label{integ_first term}
\ee 
where
\be
 k^2=\frac{k_1^2-1}{k_1^2}, \quad k_1=\sqrt{D^2-1}+D , \quad D=\frac{4r_1^2+s_1^2+s_2^2}{2s_1s_2} . 
 \label{krho_general} 
\ee
%\beqn
% && k^2=\frac{k_1^2-1}{k_1^2}, \qquad k_1=\sqrt{D^2-1}+D , \nonumber \label{krho_general} \\
% && D=\frac{4r_1^2+s_1^2+s_2^2}{2s_1s_2} .
%\eeqn
We can investigate the location of the singularity of the expression~(\ref{integ_first term}). This occurs when $k=1$, i.e. for ${s_1s_2=0}$. From (\ref{system}) one gets $s_1^2s_2^2=-3r_1^4-2Ar_1^2+C$, so that the singularity is at $r_1^2=-\frac{1}{3}(A+\sqrt{A^2+3C})$. This exactly corresponds to the explicit expression (\ref{Cardano}) for $Q=0$, i.e. $27q^2=-4p^3$. Using the relations (\ref{Qpq}) and (\ref{ABC}), it is straightforward (but somewhat lengthy) to demonstrate that this polynomial condition is satisfied for 
\be
 x_2=0, \qquad z_2^2=(L^2-y_2^2)\cos^2\alpha .
\label{source loc}
\ee
This singular behaviour of the term~(\ref{integ_first term}) suggests that there is a singular source located on the wave front (${u=0}$) of the metric~(\ref{ppgeneral}), precisely in the plane $x_2=0$ on the ellipse
\be
 \left(\frac{y_2}{L}\right)^2+\left(\frac{z_2}{L\cos\alpha}\right)^2=1 .
 \label{ellipse}
\ee
Of course the above argument is not conclusive. Rigorously, we should integrate also all the other terms in Eq.~(\ref{h_general}). This could in principle be done, but it would lead to an involved expression without much practical use. We rather prefer to integrate numerically the full function~(\ref{h_general}), and display the thus obtained profile $H$ in Fig.~\ref{fig_general}, which indeed confirms the presence of a singular ellipse within the wave front. This also corresponds to intuitive expectations, since the original static circular source has been boosted in a general direction. [Moreover, it agrees with the following argument: the source of the black ring~(\ref{ring}) was located at $y=-\infty$, i.e. $\eta=0$ and $\xi=L$. In view of Eq.~(\ref{xieta}), these conditions become Eq.~(\ref{source loc}), which is unchanged under the boost~(\ref{lorentzboost}).]

\section{Conclusions}

\label{sec_conclusions}

We have derived the gravitational field generated by a five-dimensional static black ring moving ``with the speed of light''. More precisely, we have calculated how the Emparan-Reall line element transforms under appropriate boosts, and studied the ultrarelativistic limit when the boost velocity approaches the speed of light. In particular, we have studied in detail two complementary boosts along privileged directions, namely those orthogonal and parallel to the plane containing the ring circle. The resulting line elements represent impulsive \pp waves. These are exact vacuum solutions everywhere except at singular points that are a remnant of the original curvature singularity of the static black ring. In addition, in the case of the orthogonal boost, there is a disk-shaped membrane inside the ring directly inherited from the conical singularity of the static Emparan-Reall spacetime. [Notice that the profile functions obtained via the boosting procedure ultimately provide solutions to equivalent problems of three-dimensional electrostatics (or Newtonian gravity) with a disk or a nonuniform rod source.] Further analysis of the solutions has been supplemented via graphical plots and via suitable expansions of the metric functions. We may also observe here that, if necessary, one could introduce a coordinate system in which the metric coefficients take a continuous form, using the general transformation presented in \cite{DratHo85npb}. 

It is also worth remarking that, in contradistinction to the well known situation in four dimensions \cite{AicSex71}, we did not need to perform any infinite subtractions during our calculations. This is essentially due to the faster fall-off of the gravitational potential of a ``monopole'' in $D>4$, which ensures that all the required integrals are finite. The same simplification occurred in previous investigations of ultrarelativistic boosts in higher dimensions \cite{LouSan90,Ortaggio05,Yoshino05}, as well as in the case of the boost of particles with multipole moments in $D=4$ (Weyl solutions) \cite{PodGri98prd}.  

We have concentrated on a static ring containing a disk membrane at $x=+1$, for which there is no conical singularity at infinity. A generalization to the case of a ring with a deficit membrane at $x=-1$ (which extends to infinity) would be straightforward. It would be more interesting to extend our results to the case of rotating black rings. Such work is currently in progress \cite{OrtKrtPod05_2}. 

\begin{acknowledgments}

M.O. is supported by a post-doctoral fellowship from Istituto Nazionale di Fisica Nucleare (bando n.10068/03).

\end{acknowledgments}

%\appendix*
\appendix

\section{Elliptic integrals}

\label{app_integrals}

In this Appendix we summarize the definitions and the properties of the complete elliptic integrals employed 
in the main text, following references \cite{Gradshteynbook6,GroHofbook2}. 

\subsection{Definitions}

The complete elliptic integrals in trigonometric form are defined by \cite{Gradshteynbook6}
\be
 K(k)=\int_0^{\pi/2}\frac{\d\alpha}{\sqrt{1-k^2\sin^2\alpha}} , 
\ee
\beqn 
 & & \hspace{-.3cm} E(k)=\int_0^{\pi/2}\sqrt{1-k^2\sin^2\alpha}\,\d\alpha , \\
 & & \hspace{-.3cm} \Pi(\rho,k)=\int_0^{\pi/2}\frac{\d\alpha}{(1-\rho\sin^2\alpha)\sqrt{1-k^2\sin^2\alpha}} .
\eeqn
%\beqn
% & & \hspace{-.3cm} K(k)=\int_0^{\pi/2}\frac{\d\alpha}{\sqrt{1-k^2\sin^2\alpha}} , \\
% & & \hspace{-.3cm} E(k)=\int_0^{\pi/2}\sqrt{1-k^2\sin^2\alpha}\,\d\alpha , \\
% & & \hspace{-.3cm} \Pi(\rho,k)=\int_0^{\pi/2}\frac{\d\alpha}{(1-\rho\sin^2\alpha)\sqrt{1-k^2\sin^2\alpha}} .
%\eeqn

\subsection{Useful identities}

They satisfy the identities \cite{GroHofbook2}
\be
 (k^2-\rho)\Pi(\rho,k)=k^2K(k)-\frac{\rho(1-k^2)}{1-\rho}\Pi\left(\frac{k^2-\rho}{1-\rho},k\right) , 
 \label{iden1}
\ee
\beqn
 & & \Pi(\rho,k)=K(k)-\Pi\left(\frac{k^2}{\rho},k\right)+\frac{\pi}{2}\sqrt{\frac{-\rho}{(1-\rho)(k^2-\rho)}} , \nonumber \label{iden2} \\ 
 & & \quad \mbox{with } \ \rho(1-\rho)^{-1}(k^2-\rho)^{-1}<0 .
\eeqn

\subsection{Differential relations}

Derivatives of elliptic integrals lead to combinations of the same integrals:
\beqn
 & & \frac{\d K(k)}{\d k}=\frac{E(k)}{k(1-k^2)}-\frac{K(k)}{k} , \nonumber \\
 & & \frac{\d E(k)}{\d k}=\frac{E(k)-K(k)}{k} ,  \nonumber \\
 & & \frac{\pa\Pi(\rho,k)}{\pa k}=\frac{k}{k^2-\rho}\left[-\Pi(\rho,k)+\frac{E(k)}{1-k^2}\right] , \label{differential} \\
 & & \frac{\pa\Pi(\rho,k)}{\pa \rho}=\frac{1}{2\rho(1-\rho)} \nonumber \\ 
 & & \qquad {}\times\left[\frac{k^2-\rho^2}{k^2-\rho}\Pi(\rho,k)-K(k)-\frac{\rho}{k^2-\rho}E(k)\right] .  \nonumber
\eeqn 

\subsection{Series representations}

The behaviour near $k=0$ is given by
\be
 K(k)=\frac{\pi}{2}\left(1+\frac{1}{4}k^2+\frac{9}{64}k^4+O(k^6)\right) , 
 \label{serK1}
\ee
\be
 E(k)=\frac{\pi}{2}\left(1-\frac{1}{4}k^2-\frac{3}{64}k^4+O(k^6)\right) , 
 \label{serE1}
\ee
\beqn
 & & \Pi(\rho,k)= \frac{\pi}{2}\sum_{\mu=0}^\infty\sum_{\nu=0}^\mu\frac{(2\mu-1)!!(2\nu-1)!!}{(2\mu)!!(2\nu)!!}k^{2\nu}\rho^{\mu-\nu}
 , \label{serPi1}  \nonumber \\
 & & \quad \mbox{with } \ |\rho|<1 .
\eeqn
%\beqn
% & & \Pi(\rho,k)= \frac{\pi}{2}\sum_{\mu=0}^\infty\sum_{\nu=0}^\mu\frac{(1;2;\mu)(1;2;\nu)}{(2;2;\mu)(2;2;\nu)}k^{2\nu}\rho^{\mu-\nu}
% , \label{serPi1} \\
% & & \mbox{with } \ |\rho|<1 \ \mbox{ and the definitions } \nonumber \\
% & & (m;n;0)=1 , \nonumber \\
% & & (m;n;p)=m(m+n)(m+2n)\ldots (m+(p-1)n) . \nonumber 
%\eeqn

Near the singular point $k=1$ one has
\beqn
 & & K(k)=-\frac{1}{2}\log\frac{1-k^2}{16}-\frac{1}{8}\left(2+\log\frac{1-k^2}{16}\right)(1-k^2) \nonumber \\
 & & \qquad \qquad {}-\frac{9}{128}\left(\frac{7}{3}+\log\frac{1-k^2}{16}\right)(1-k^2)^2 \nonumber  \label{serK2} \\ 
 & & \qquad \qquad {}+O\left((1-k^2)^3\right) , \\
 & & E(k)=1-\frac{1}{4}\left(1+\log\frac{1-k^2}{16}\right)(1-k^2) \nonumber \\
 & & \qquad \qquad {}-\frac{3}{32}\left(\frac{13}{6}+\log\frac{1-k^2}{16}\right)(1-k^2)^2\nonumber  \label{serE2} \\ 
 & & \qquad \qquad {}+O\left((1-k^2)^3\right) , \\
 & & \Pi(\rho,k)=\frac{1}{1-\rho}\log\frac{4}{\sqrt{1-k^2}}+\frac{\sqrt{-\rho}}{1-\rho}\arctan\sqrt{-\rho} \nonumber \label{serPi2} \\ 
 & & \qquad \qquad {}+O(1-k^2) , \qquad \mbox{with } \ \rho<0 . 
\eeqn

\subsection{Useful integrals: orthogonal boost}

In Sec.~\ref{sec_orthogonal} we employed the following integrals:
\beqn
 & & \hspace{-.3cm} \int_0^{\infty}\frac{\d z}{\sqrt{(z^2+a^2)(z^2+b^2)}}=\frac{1}{a}K(k) , \label{orth1} \\
 & & \hspace{-.3cm} \int_0^{\infty}\frac{z^2\d z}{\sqrt{(z^2+a^2)^3(z^2+b^2)^3}}= \nonumber \\ \label{orth2}
 & & \quad \frac{a^2+b^2}{a(a^2-b^2)^2}K(k)-\frac{2a}{(a^2-b^2)^2}E(k) , \\
 & & \hspace{-.3cm} \int_0^{\infty}\frac{\d z}{(z^2+c^2)\sqrt{(z^2+a^2)(z^2+b^2)}}= \nonumber  \label{orth3} \\ 
 & & \quad \hspace{-.4cm} \frac{1}{a(b^2-c^2)}\left[\frac{b^2}{c^2}\Pi\left(-\frac{b^2-c^2}{c^2},k\right)-K(k)\right] ,
\eeqn
where
\be
 k=\frac{\sqrt{a^2-b^2}}{a} , \quad a>b>0, \quad c\neq 0 . 
\ee

\subsection{Useful integrals: parallel boost}

The integrals used in Sec.~\ref{sec_parallel} are:
\beqn
 & & \int_0^{\infty}\frac{\d z}{\sqrt{z^4+2b^2z^2+a^4}}=\frac{1}{a}K(k) , \label{parall1} \\
 & & \int_0^{\infty}\frac{z^2\d z}{\sqrt{(z^4+2b^2z^2+a^4)^3}}= \nonumber   \label{parall2} \\
 & & \qquad \frac{a}{a^4-b^4}E(k)-\frac{1}{2a(a^2-b^2)}K(k) , \\
 & & \int_0^{\infty}\frac{\d z}{(z^2+c^2)\sqrt{z^4+2b^2z^2+a^4}}= \label{parall3} \\ 
 & & \quad \frac{1}{a(a^2-c^2)}\left[\frac{a^2+c^2}{2c^2}\Pi\left(-\frac{(a^2-c^2)^2}{4c^2a^2},k\right)-K(k)\right] ,  \nonumber
\eeqn
with
\be
 k=\frac{\sqrt{a^2-b^2}}{\sqrt{2}a} , \quad a^2>b^2>-\infty , \quad a^2>0, \quad c\neq 0 . 
\ee

\section{The Weyl tensor for $H_{_{\!\bot}}$}

\label{app_weyl}

Here we present explicitly the frame components of the Weyl tensor in the case of the metric (\ref{pporth}), (\ref{Horth}) describing a black ring boosted in an orthogonal direction. Using Eq.~(\ref{differential}), from Eq.~(\ref{Weylscal}) with Eq.~(\ref{Horth}) we obtain
\beqn
  & & \Psi_{11}=\frac{2\sqrt{2}p}{3\pi L^2}\frac{1}{\xi^2\sqrt{[(\xi+L)^2+x_2^2]^3}} \nonumber \\ 
  & & \quad {}\times\Bigg\{\!-\left[L^6+(x_2^2-2\xi^2)L^4+(\xi^4-x_2^4+8\xi^2x_2^2)L^2\right. \nonumber \\ 
  & & \qquad\qquad {}-\left.x_2^2(x_2^2+\xi^2)^2\right]\frac{K(k)}{(\xi-L)^2+x_2^2} \nonumber \\
  & & \quad {}+\left[L^8+(2x_2^2-7\xi^2)L^6+\xi^2(11\xi^2+7x_2^2)L^4\right. \nonumber \\
  & & \quad {}-(x_2^2+\xi^2)(2x_2^4-13\xi^2x_2^2+5\xi^4)L^2-\left.x_2^2(x_2^2+\xi^2)^3\right] \nonumber \\
 & & \qquad\qquad {}\times\frac{E(k)}{[(\xi-L)^2+x_2^2]^2}\Bigg\}\delta(u)-\frac{1}{6}R_{uu} ,  
\eeqn
\beqn
 & & \Psi_{13}=\frac{2\sqrt{2}p}{3\pi L^2}\frac{x_2}{\xi\sqrt{[(\xi+L)^2+x_2^2]^3}} \nonumber \\
 & & \quad {}\times\Bigg\{\!-\frac{3L^4+4(x_2^2-\xi^2)L^2+(\xi^2+x_2^2)^2}{(\xi-L)^2+x_2^2}K(k) \nonumber \\ 
 & & \quad\quad {}+\left[3L^6+(11\xi^2+7x_2^2)L^4+5(x_2^2-3\xi^2)(x_2^2+\xi^2)L^2\right. \nonumber \\
 & & \quad\quad  {}+\left.(x_2^2+\xi^2)^3\right]\frac{E(k)}{[(\xi-L)^2+x_2^2]^2}\Bigg\}\delta(u) , 
\eeqn
\beqn
 & & \Psi_{22}=\frac{4\sqrt{2}p}{3\pi L^2}\frac{1}{\xi\sqrt{(\xi+L)^2+x_2^2}} \nonumber \\
 & & \quad {}\times\left\{(L^2-\xi^2-x_2^2)K(k)-\frac{L^4-(\xi^2+x_2^2)^2}{(\xi-L)^2+x_2^2}E(k)\right\}\delta(u)     
    \nonumber \\
 & & \quad {}-\frac{1}{6}R_{uu} ,
\eeqn
\be
 \Psi_{33}=-(\Psi_{11}+\Psi_{22}) .
\ee
The last equation follows from the tracelessness of the Weyl tensor.

%\bibliography{bibl}

\begin{thebibliography}{28}
\expandafter\ifx\csname natexlab\endcsname\relax\def\natexlab#1{#1}\fi
\expandafter\ifx\csname bibnamefont\endcsname\relax
  \def\bibnamefont#1{#1}\fi
\expandafter\ifx\csname bibfnamefont\endcsname\relax
  \def\bibfnamefont#1{#1}\fi
\expandafter\ifx\csname citenamefont\endcsname\relax
  \def\citenamefont#1{#1}\fi
\expandafter\ifx\csname url\endcsname\relax
  \def\url#1{\texttt{#1}}\fi
\expandafter\ifx\csname urlprefix\endcsname\relax\def\urlprefix{URL }\fi
\providecommand{\bibinfo}[2]{#2}
\providecommand{\eprint}[2][]{\url{#2}}

\bibitem[{\citenamefont{Myers and Perry}(1986)}]{MyePer86}
\bibinfo{author}{\bibfnamefont{R.~C.} \bibnamefont{Myers}} \bibnamefont{and}
  \bibinfo{author}{\bibfnamefont{M.~J.} \bibnamefont{Perry}},
  \bibinfo{journal}{Ann. Phys. (N.Y.)} \textbf{\bibinfo{volume}{172}},
  \bibinfo{pages}{304} (\bibinfo{year}{1986}).

\bibitem[{\citenamefont{Banks and Fischler}()}]{BanFis99}
\bibinfo{author}{\bibfnamefont{T.}~\bibnamefont{Banks}} \bibnamefont{and}
  \bibinfo{author}{\bibfnamefont{W.}~\bibnamefont{Fischler}},
  \bibinfo{note}{{``A model for high energy scattering in quantum gravity''}
  [hep-th/9906038]}.

\bibitem[{\citenamefont{Emparan et~al.}(2000)\citenamefont{Emparan, Horowitz,
  and Myers}}]{EmpHorMye00prl}
\bibinfo{author}{\bibfnamefont{R.}~\bibnamefont{Emparan}},
  \bibinfo{author}{\bibfnamefont{G.~T.} \bibnamefont{Horowitz}},
  \bibnamefont{and} \bibinfo{author}{\bibfnamefont{R.~C.} \bibnamefont{Myers}},
  \bibinfo{journal}{Phys. Rev. Lett.} \textbf{\bibinfo{volume}{85}},
  \bibinfo{pages}{499} (\bibinfo{year}{2000}).

\bibitem[{\citenamefont{Giddings and Thomas}(2002)}]{GidTho02}
\bibinfo{author}{\bibfnamefont{S.~B.} \bibnamefont{Giddings}} \bibnamefont{and}
  \bibinfo{author}{\bibfnamefont{S.}~\bibnamefont{Thomas}},
  \bibinfo{journal}{Phys. Rev. {\rm D}} \textbf{\bibinfo{volume}{65}},
  \bibinfo{pages}{056010} (\bibinfo{year}{2002}).

\bibitem[{\citenamefont{Dimopoulos and Landsberg}(2001)}]{DimLan01}
\bibinfo{author}{\bibfnamefont{S.}~\bibnamefont{Dimopoulos}} \bibnamefont{and}
  \bibinfo{author}{\bibfnamefont{G.}~\bibnamefont{Landsberg}},
  \bibinfo{journal}{Phys. Rev. Lett.} \textbf{\bibinfo{volume}{87}},
  \bibinfo{pages}{161602} (\bibinfo{year}{2001}).

\bibitem[{\citenamefont{Dray and 't~Hooft}(1985)}]{DratHo85npb}
\bibinfo{author}{\bibfnamefont{T.}~\bibnamefont{Dray}} \bibnamefont{and}
  \bibinfo{author}{\bibfnamefont{G.}~\bibnamefont{'t~Hooft}},
  \bibinfo{journal}{Nucl. Phys. {\rm B}} \textbf{\bibinfo{volume}{253}},
  \bibinfo{pages}{173} (\bibinfo{year}{1985}).

\bibitem[{\citenamefont{'t~Hooft}(1987)}]{tHooft87}
\bibinfo{author}{\bibfnamefont{G.}~\bibnamefont{'t~Hooft}},
  \bibinfo{journal}{Phys. Lett. {\rm B}} \textbf{\bibinfo{volume}{198}},
  \bibinfo{pages}{61} (\bibinfo{year}{1987}).

\bibitem[{\citenamefont{Aichelburg and Sexl}(1971)}]{AicSex71}
\bibinfo{author}{\bibfnamefont{P.~C.} \bibnamefont{Aichelburg}}
  \bibnamefont{and} \bibinfo{author}{\bibfnamefont{R.~U.} \bibnamefont{Sexl}},
  \bibinfo{journal}{Gen. Rel. Grav.} \textbf{\bibinfo{volume}{2}},
  \bibinfo{pages}{303} (\bibinfo{year}{1971}).

\bibitem[{\citenamefont{Loust\'o and S\'anchez}(1990)}]{LouSan90}
\bibinfo{author}{\bibfnamefont{C.~O.} \bibnamefont{Loust\'o}} \bibnamefont{and}
  \bibinfo{author}{\bibfnamefont{N.}~\bibnamefont{S\'anchez}},
  \bibinfo{journal}{Int. J. Mod. Phys. {\rm A}} \textbf{\bibinfo{volume}{5}},
  \bibinfo{pages}{915} (\bibinfo{year}{1990}).

\bibitem[{\citenamefont{Ortaggio}(2005)}]{Ortaggio05}
\bibinfo{author}{\bibfnamefont{M.}~\bibnamefont{Ortaggio}},
  \bibinfo{journal}{JHEP} \textbf{\bibinfo{volume}{05}}, \bibinfo{pages}{048}
  (\bibinfo{year}{2005}).

\bibitem[{\citenamefont{Eardley and Giddings}(2002)}]{EarGid02}
\bibinfo{author}{\bibfnamefont{D.~M.} \bibnamefont{Eardley}} \bibnamefont{and}
  \bibinfo{author}{\bibfnamefont{S.~B.} \bibnamefont{Giddings}},
  \bibinfo{journal}{Phys. Rev. {\rm D}} \textbf{\bibinfo{volume}{66}},
  \bibinfo{pages}{044011} (\bibinfo{year}{2002}).

\bibitem[{\citenamefont{Kohlprath and Veneziano}(2002)}]{KohVen02}
\bibinfo{author}{\bibfnamefont{E.}~\bibnamefont{Kohlprath}} \bibnamefont{and}
  \bibinfo{author}{\bibfnamefont{G.}~\bibnamefont{Veneziano}},
  \bibinfo{journal}{JHEP} \textbf{\bibinfo{volume}{06}}, \bibinfo{pages}{057}
  (\bibinfo{year}{2002}).

\bibitem[{\citenamefont{Yoshino and Nambu}(2002)}]{YosNam02}
\bibinfo{author}{\bibfnamefont{H.}~\bibnamefont{Yoshino}} \bibnamefont{and}
  \bibinfo{author}{\bibfnamefont{Y.}~\bibnamefont{Nambu}},
  \bibinfo{journal}{Phys. Rev. {\rm D}} \textbf{\bibinfo{volume}{66}},
  \bibinfo{pages}{065004} (\bibinfo{year}{2002}).

\bibitem[{\citenamefont{Giddings and Rychkov}(2004)}]{GidRyc04}
\bibinfo{author}{\bibfnamefont{S.~B.} \bibnamefont{Giddings}} \bibnamefont{and}
  \bibinfo{author}{\bibfnamefont{V.~S.} \bibnamefont{Rychkov}},
  \bibinfo{journal}{Phys. Rev. {\rm D}} \textbf{\bibinfo{volume}{70}},
  \bibinfo{pages}{104026} (\bibinfo{year}{2004}).

\bibitem[{\citenamefont{Yoshino}(2005)}]{Yoshino05}
\bibinfo{author}{\bibfnamefont{H.}~\bibnamefont{Yoshino}},
  \bibinfo{journal}{Phys. Rev. {\rm D}} \textbf{\bibinfo{volume}{71}},
  \bibinfo{pages}{044032} (\bibinfo{year}{2005}).

\bibitem[{\citenamefont{Emparan and Reall}(2002{\natexlab{a}})}]{EmpRea02prl}
\bibinfo{author}{\bibfnamefont{R.}~\bibnamefont{Emparan}} \bibnamefont{and}
  \bibinfo{author}{\bibfnamefont{H.~S.} \bibnamefont{Reall}},
  \bibinfo{journal}{Phys. Rev. Lett.} \textbf{\bibinfo{volume}{88}},
  \bibinfo{pages}{101101} (\bibinfo{year}{2002}{\natexlab{a}}).

\bibitem[{\citenamefont{Emparan and Reall}(2002{\natexlab{b}})}]{EmpRea02prd}
\bibinfo{author}{\bibfnamefont{R.}~\bibnamefont{Emparan}} \bibnamefont{and}
  \bibinfo{author}{\bibfnamefont{H.~S.} \bibnamefont{Reall}},
  \bibinfo{journal}{Phys. Rev. {\rm D}} \textbf{\bibinfo{volume}{65}},
  \bibinfo{pages}{084025} (\bibinfo{year}{2002}{\natexlab{b}}).

\bibitem[{\citenamefont{Ortaggio et~al.}(2005)\citenamefont{Ortaggio, Krtou{\v
  s}, and Podolsk\'y}}]{OrtKrtPod05_2}
\bibinfo{author}{\bibfnamefont{M.}~\bibnamefont{Ortaggio}},
  \bibinfo{author}{\bibfnamefont{P.}~\bibnamefont{Krtou{\v s}}},
  \bibnamefont{and}
  \bibinfo{author}{\bibfnamefont{J.}~\bibnamefont{Podolsk\'y}}
  (\bibinfo{year}{2005}), \bibinfo{note}{[gr-qc/0506064]}.

\bibitem[{\citenamefont{Emparan}(2004)}]{Emparan04}
\bibinfo{author}{\bibfnamefont{R.}~\bibnamefont{Emparan}},
  \bibinfo{journal}{JHEP} \textbf{\bibinfo{volume}{03}}, \bibinfo{pages}{064}
  (\bibinfo{year}{2004}).

\bibitem[{\citenamefont{Pravda and Pravdov\'a}(2005)}]{PraPra05}
\bibinfo{author}{\bibfnamefont{V.}~\bibnamefont{Pravda}} \bibnamefont{and}
  \bibinfo{author}{\bibfnamefont{A.}~\bibnamefont{Pravdov\'a}}
  (\bibinfo{year}{2005}), \bibinfo{note}{to appear in Gen. Rel. Grav.
  [gr-qc/0501003]}.

\bibitem[{\citenamefont{Coley et~al.}(2004)\citenamefont{Coley, Milson, Pravda,
  and Pravdov\'a}}]{Coleyetal04}
\bibinfo{author}{\bibfnamefont{A.}~\bibnamefont{Coley}},
  \bibinfo{author}{\bibfnamefont{R.}~\bibnamefont{Milson}},
  \bibinfo{author}{\bibfnamefont{V.}~\bibnamefont{Pravda}}, \bibnamefont{and}
  \bibinfo{author}{\bibfnamefont{A.}~\bibnamefont{Pravdov\'a}},
  \bibinfo{journal}{Class. Quantum Grav.} \textbf{\bibinfo{volume}{21}},
  \bibinfo{pages}{L35} (\bibinfo{year}{2004}).

\bibitem[{\citenamefont{Aichelburg and Balasin}(1996)}]{AicBal96}
\bibinfo{author}{\bibfnamefont{P.~C.} \bibnamefont{Aichelburg}}
  \bibnamefont{and} \bibinfo{author}{\bibfnamefont{H.}~\bibnamefont{Balasin}},
  \bibinfo{journal}{Class. Quantum Grav.} \textbf{\bibinfo{volume}{13}},
  \bibinfo{pages}{723} (\bibinfo{year}{1996}).

\bibitem[{\citenamefont{Podolsk\'y and Ortaggio}(2001)}]{PodOrt01}
\bibinfo{author}{\bibfnamefont{J.}~\bibnamefont{Podolsk\'y}} \bibnamefont{and}
  \bibinfo{author}{\bibfnamefont{M.}~\bibnamefont{Ortaggio}},
  \bibinfo{journal}{Class. Quantum Grav.} \textbf{\bibinfo{volume}{18}},
  \bibinfo{pages}{2689} (\bibinfo{year}{2001}).

\bibitem[{\citenamefont{Loust\'o and S\'anchez}(1991)}]{LouSan91}
\bibinfo{author}{\bibfnamefont{C.~O.} \bibnamefont{Loust\'o}} \bibnamefont{and}
  \bibinfo{author}{\bibfnamefont{N.}~\bibnamefont{S\'anchez}},
  \bibinfo{journal}{Nucl. Phys. {\rm B}} \textbf{\bibinfo{volume}{355}},
  \bibinfo{pages}{231} (\bibinfo{year}{1991}).

\bibitem[{\citenamefont{Pravda et~al.}(2004)\citenamefont{Pravda, Pravdov\'a,
  Coley, and Milson}}]{Pravdaetal04}
\bibinfo{author}{\bibfnamefont{V.}~\bibnamefont{Pravda}},
  \bibinfo{author}{\bibfnamefont{A.}~\bibnamefont{Pravdov\'a}},
  \bibinfo{author}{\bibfnamefont{A.}~\bibnamefont{Coley}}, \bibnamefont{and}
  \bibinfo{author}{\bibfnamefont{R.}~\bibnamefont{Milson}},
  \bibinfo{journal}{Class. Quantum Grav.} \textbf{\bibinfo{volume}{21}},
  \bibinfo{pages}{2873} (\bibinfo{year}{2004}).

\bibitem[{\citenamefont{Gradshteyn and Ryzhik}(2000)}]{Gradshteynbook6}
\bibinfo{author}{\bibfnamefont{I.~S.} \bibnamefont{Gradshteyn}}
  \bibnamefont{and} \bibinfo{author}{\bibfnamefont{I.~M.}
  \bibnamefont{Ryzhik}}, \emph{\bibinfo{title}{Table of Integrals, Series, and
  Products}} (\bibinfo{publisher}{Academic Press}, \bibinfo{address}{San
  Diego}, \bibinfo{year}{2000}), \bibinfo{edition}{sixth} ed.

\bibitem[{\citenamefont{Gr{\"{o}}bner and Hofreiter}(1966)}]{GroHofbook2}
\bibinfo{author}{\bibfnamefont{W.}~\bibnamefont{Gr{\"{o}}bner}}
  \bibnamefont{and}
  \bibinfo{author}{\bibfnamefont{N.}~\bibnamefont{Hofreiter}},
  \emph{\bibinfo{title}{Integraltafel}}, vol.~\bibinfo{volume}{2}
  (\bibinfo{publisher}{Springer-Verlag}, \bibinfo{address}{Wien},
  \bibinfo{year}{1966}).

\bibitem[{\citenamefont{Podolsk\'y and Griffiths}(1998)}]{PodGri98prd}
\bibinfo{author}{\bibfnamefont{J.}~\bibnamefont{Podolsk\'y}} \bibnamefont{and}
  \bibinfo{author}{\bibfnamefont{J.~B.} \bibnamefont{Griffiths}},
  \bibinfo{journal}{Phys. Rev. {\rm D}} \textbf{\bibinfo{volume}{58}},
  \bibinfo{pages}{124024} (\bibinfo{year}{1998}).

\end{thebibliography}

\end{document}